# Quality control tests of the front-end optical link components for the ATLAS Liquid Argon Calorimeter Phase-1 upgrade

B. Deng,[a,b] J. Thomas,[c] L. Zhang,[c,d,1] E. Baker,[e] A. Barsallo,[c] M.L. Bleile,[c]
C. Chen,[c,d] I. Cohen,[c] E. Cruda,[c] J. Fang,[c] N. Feng,[c] D. Gong,[c] S. Hou,[f] X. Huang,[c,d]
T. Lozano-Brown,[c] C. Liu,[c] T. Liu,[c,*] A. Muhammad,[c] L.A. Murphy,[c] P.M. Price,[c] J.H.
Ray,[c] C. Rhoades,[c] A.H. Santhi,[c] D. Sela,[g,h] H. Sun,[c,d] J. Wang,[i] Z. Wang,[c] Z. Wang,[i]
L. Xiao,[c,d] W. Zhang,[c,d] X. Zhao,[c] and J. Ye[c]

[a] *Hubei Polytechnic University,*
  *Huangshi, Hubei 435003, China*
[b] *Zhongyi Environmental Protection College,*
  *Yixing, Jiangsu 214200, China*
[c] *Southern Methodist University,*
  *Dallas, TX 75275, USA*
[d] *Central China Normal University,*
  *Wuhan, Hubei 430079, China*
[e] *Keller High School,*
  *Keller, TX 76248, USA*
[f] *Academia Sinica,*
  *Nangang, Taipei 11529, Taiwan*
[g] *Yorktown Education,*
  *Plano, TX 75093, USA*
[h] *Collin College,*
  *McKinney, TX 75071, USA*
[i] *University of Science and Technology of China*
  *Hefei, Anhui 230026, China*
[1] *Visiting scholar at SMU*
  *E-mail:* tliu@smu.edu

ABSTRACT: We present the procedures and results of the quality control tests for the front-end optical link components in the ATLAS Liquid Argon Calorimeter Phase-1 upgrade. The components include a Vertical-Cavity Surface-Emitting Laser (VCSEL) driver ASIC LOCld, custom optical transmitter/transceiver modules MTx/MTRx, and a transmitter ASIC LOCx2. LOCld, MTx, and LOCx2 each contain two channels with the same structure, while MTRx has a transmitter channel and a receiver channel. Each channel is tested at 5.12 Gbps. A total of 5341 LOCld chips, 3275 MTx modules, 797 MTRx modules, and 3198 LOCx2 chips are qualified. The yields are 73.9%, 98.0%, 98.4%, and 61.9% for LOCld, LOCx2, MTx, and MTRx, respectively.

---

[*] Corresponding author.

# Contents

## 1. Introduction

The ATLAS detector at the LHC has been operating for over a decade [1, 2]. As the Large Hadron Collider progresses towards higher luminosity, the ATLAS detector is being upgraded to maintain the selectivity of the online trigger in the presence of a much greater number of events in every beam crossing. The trigger readout electronics of the Liquid Argon (LAr) Calorimeter is one part of the Phase-1 upgrade to the ATLAS detector [3].

The block diagram of the upgraded trigger readout electronics of the ATLAS LAr Calorimeter is shown in figure 1. The detector signals are transmitted from the detector to the counting room via upward data optical links. As part of the data optical links, a dual-channel transmitter ASIC called LOCx2 [4] encodes and serializes the parallel digitized data into serial data. A dual-channel driver ASIC LOCld [5] drives two Vertical-Cavity Surface-Emitting Laser (VCSEL) diodes that convert the serial data from electrical signals to optical signals. LOCld and LOCx2 are fabricated in a 250-nm Silicon on Sapphire CMOS technology. LOCld is assembled in a dual-channel Miniature optical Transmitter (MTx) module [6].

Besides the data optical links from the detector to the counting room, we also need bidirectional optical links to provide clocks and control signals for the front-end electronics [7]. On the one hand, the front-end electronics recovers clocks (red lines in figure 1) and Bunch

Crossing Reset (BCR) signals (blue lines in figure 1) and is configured and controlled through the downward optical links data. On the other hand, the front-end electronics transmits the operation status upward to the counting room for monitoring. The control and the operation status are represented as slow control and monitor (green lines) in figure 1. A Miniature optical Transceiver (MTRx) module provides the optical interface of the control links. The MTRx module employs a LOCld on the transmitter side and a GBTIA-embedded Receiver Optical Sub-Assembly (ROSA) [8] on the receiver side. A Giga-Bit Transceiver (GBTX) [9] is the serializer/deserializer and data encoder/decoder of the control links and provides the clock and the BCR signals for LOCx2 and ADCs. A GBT Slow Control Adapter (SCA) [10] is adopted to configure LOCx2 and LOCld inside MTx/MTRx via the Inter-Integrated Circuit ($I^2C$) interface and monitor the operation status.

The height of MTx/MTRx is limited to 6 mm to meet the mechanical constraints of the Liquid Argon Trigger Digitizing Boards (LTDBs) [11]. When assembled in MTRx, only one channel of LOCld is powered on and operates at 4.8 Gbps. When assembled in MTx, both channels of LOCld are powered on, and each operates at 5.12 Gbps. Each channel of LOCx2 operates at 5.12 Gbps.

In the ATLAS LAr calorimeter Phase-1 upgrade, 150 LTDBs will be installed. Each board will contain 20 MTx modules, 5 MTRx modules, and 20 LOCx2 chips. A total of 3000 MTx modules, 750 MTRx modules, and 3000 LOCx2 will be installed, excluding extra components as spares.

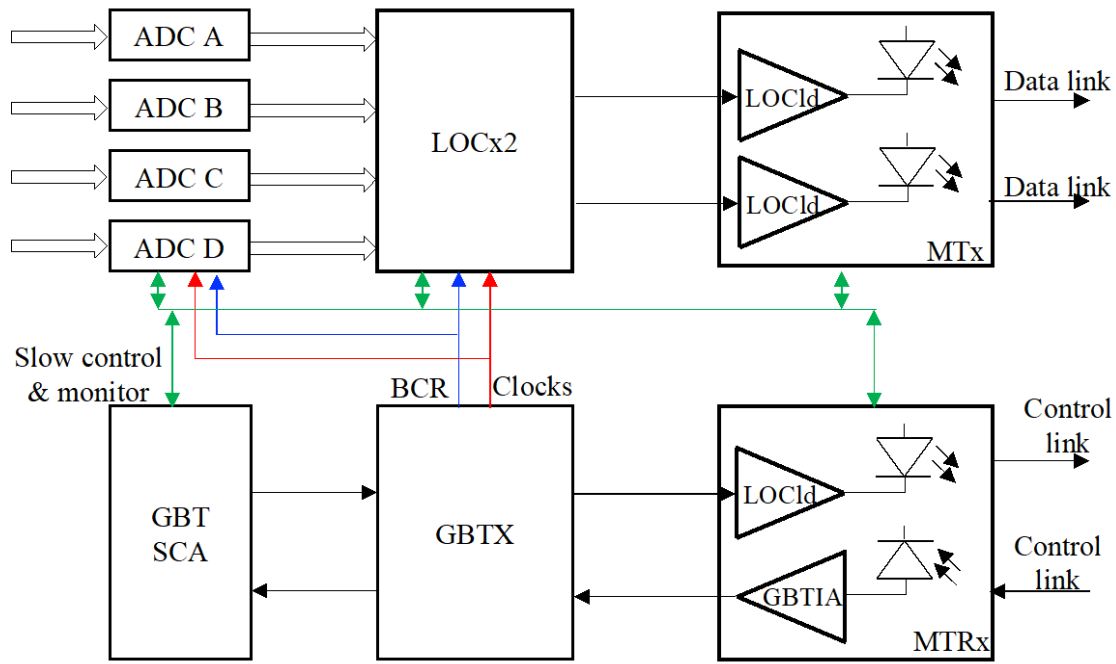

Figure 1. Block diagram of the optical link system for the ATLAS LAr calorimeter upgraded trigger readout system.

The main purpose of the Quality Control (QC) tests is to ensure that a component meets the functional requirements before it is installed to the LTDB. The screening is not a full-scale characterization. The QC tests of LOCld and LOCx2 are performed after the dies are packaged into chips. The QC test of MTx/MTRx is carried out at the module level. The QC tests allow us to remove any components with issues caused by fabrication, packaging, or assembly and reduce the cost of reworking the large LTDBs.

The primary test of LOCld, LOCx2, MTx, and MTRx is a measurement of the eye diagram. A template is used to define a sufficiently open eye. For LOCx2, a bit error rate (BER) test is also performed for each device, and a cut is applied (as detailed in section 4.2).

## 2. Quality Control of LOCld

### 2.1 Overview of LOCld

The block diagram of LOCld is shown in figure 2(a). The input of LOCld is differential Current Mode Logic (CML) with a minimum swing of 100 mV over the internal 100 Ω termination. Each channel of LOCld includes seven stages of differential amplifiers to provide high bandwidth and large gain to drive a VSCEL. The amplifiers implement active shunt-peaking inductors instead of on-chip passive inductors to save chip area. LOCld has a five-bit voltage Digital-to-Analog Converter (DAC) for the peaking voltage adjustment and two four-bit current DACs for the modulation current and the bias current. The modulation current is programmable from 7.8 to 10.6 mA, while the bias current is programmable from 2 to 15 mA. The two driving channels share a single I$^2$C target functional block. The DACs and the I$^2$C target operate at 2.5V, while all amplification stages work at 3.3V. The output differential signal of LOCld is CML with 100 Ω output resistance.

The dimension of the LOCld die is 1.090 mm by 2.114 mm. LOCld is packaged in a 40-pin Quad Frameless (QFN) plastic package with a pin pitch of 0.5 mm and a dimension of 6.0 mm by 6.0 mm. A photograph of LOCld chips on a storage tray with two flipped chips is shown in figure 2(b).

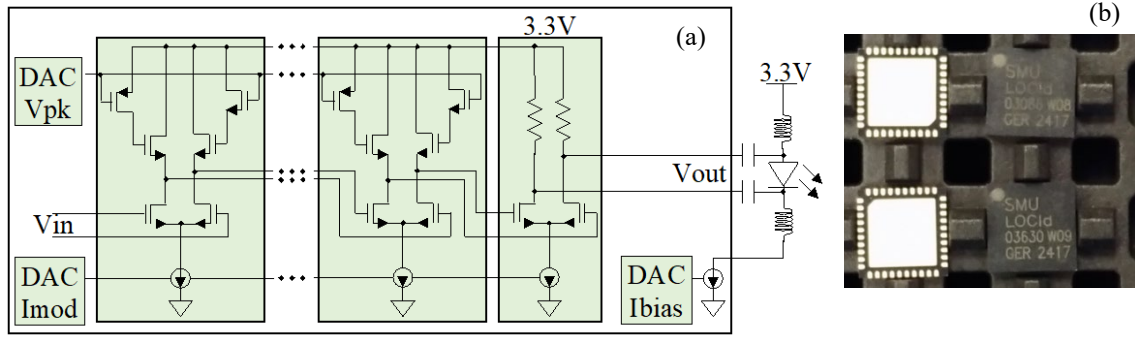

Figure 2. Block diagram of each channel of LOCld (a) and photograph of packaged LOCld chips on a storage tray.

### 2.2 Procedure of LOCld QC test

The QC test procedure of LOCld is as follows. First, each chip was placed in a socket on a test board, and the power supplies were turned on with the maximum current limited to twice the nominal values. If the supply currents reached the limits, the chip was removed as a short failure. Second, an eye diagram was recorded using the default chip configuration and compared to a mask. Any chip that failed the eye mask test was marked as eye mask failure. Third, the chip was configured using the I$^2$C interface. During this step, four-byte preset values were written to the LOCld registers, and then the registers were read back. If the read values did not match the written values, the write-read process was attempted again up to a total of 20 times, but the maximum allowed number of the I$^2$C operations was eventually set to be ten, which will be further discussed in Section 2.4. Fourth, the bias current was measured for each channel. If the bias current was more than 25% away from the expected value, the chip was classified as a bias current failure. Fifth, the supply current was measured for each channel. If any power supply consumption was more than three times the standard deviations away from the mean value, the chip was classified as supply current failure. Sixth, eye diagrams were collected for each channel. Finally, each failed chip was double-checked in another socket. Table 1 summarizes the qualification criteria of the QC test.

Table 1. Qualification criteria of the LOCld QC test.

| Parameters | Min | Max |
|---|---|---|
| Points falling in any mask | | 0 |
| Number of the I²C operations | 1 | 10 |
| Bias current | | +25% |
| Supply currents away from the mean value | $-3\sigma$ | $+3\sigma$ |

## 2.3 Setup of LOCld QC test

The block diagram is shown in figure 3(a). A pattern generator (Centellax, model number PCB12500 with a clock synthesizer TG1C1-A) was used to generate a Pseudo-Random Binary Sequence (PRBS) pattern $2^7-1$. A dedicated QC board was designed for the LOCld QC test. A fanout (Onsemi, part number NB7VQ14) was implemented to split the single pattern source into the two separate chips under test on the QC board. Passive attenuators were realized, and LOCld was tested with the minimum input swing. Two sockets (Plastronics, part number 40QHC50Y36060) were installed on the QC test board to hold the chips under test. A high-speed real-time oscilloscope (Tektronix, model number DSA72004B with two P7380 SMA differential probes) was utilized to observe eye diagrams. A USB-to-I²C adapter (Robot Electronics USB-ISS Multifunction USB Communication Module) was utilized to configure the chips under test. A laptop computer running LabVIEW controlled the test procedure. The supply currents and the bias current were monitored using a current monitor chip (Texas Instruments, part number LTC2991). The pictures of the test setup and the test board are shown in figure 3(b-c).

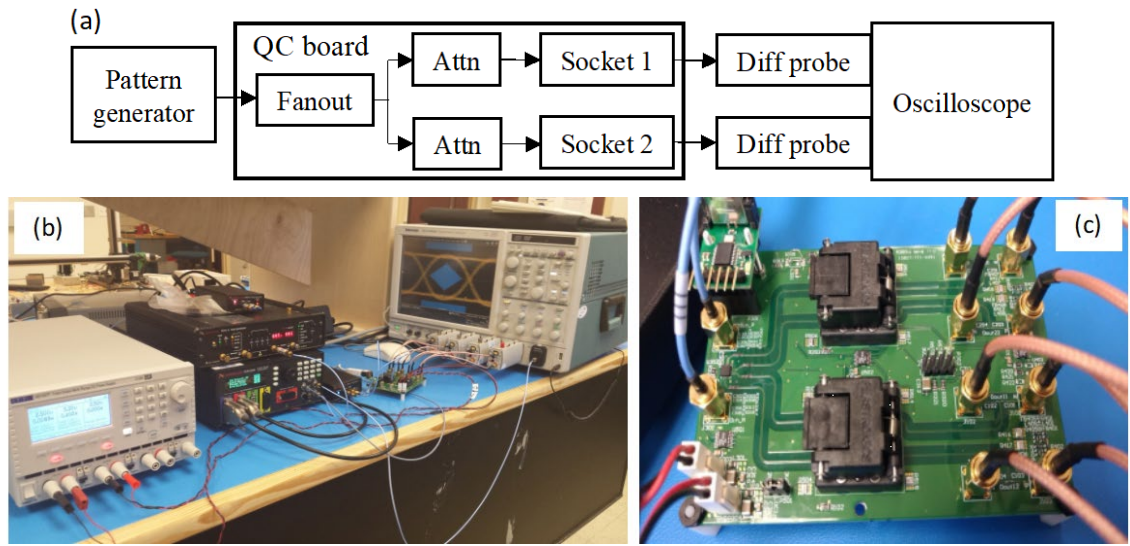

Figure 3. Block diagram of the test setup (a) and photographs of the test setup (b) and the QC board (c).

## 2.4 Results of LOCld QC test

During the QC procedure, eye diagrams were collected before (with the default power-on configuration) and after each chip was configured. Typical eye diagrams are shown in figure 4. A custom eye mask was created for this test based on the VTRx standard [12]. It can be seen that the configuration increases the peaking voltage of the waveform, which in turn affects the amplitude of the eye diagram. Previous testing has shown that the overshoot will decrease as the circuit accumulates the total ionizing dose.

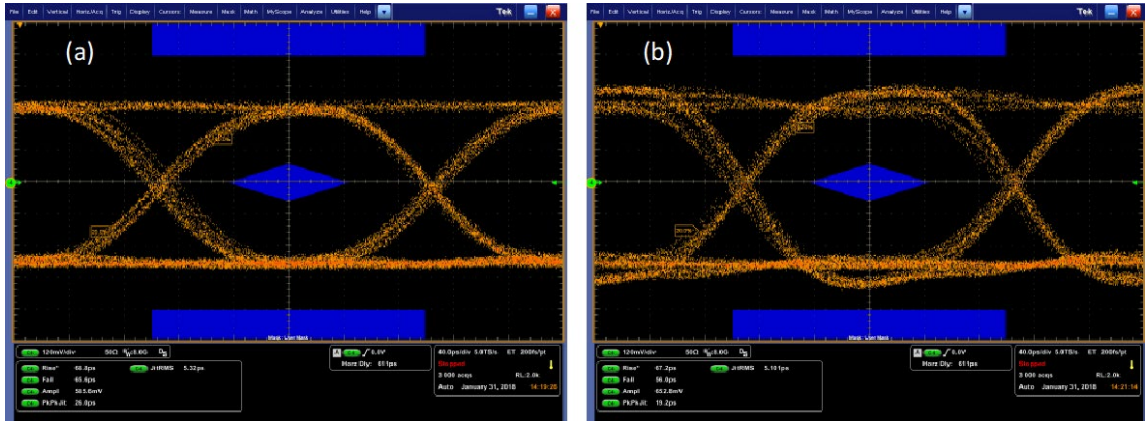

Figure 4. Typical eye diagrams without configuration (a) and after configuration (b).

The QC test yield and the percentage of each failure mode in the LOCld QC test are shown in figure 5. A total of 7227 LOCld chips were produced and packaged, and 5341 chips were fully qualified. The overall yield is 73.9%.

The failure modes of the QC test are distributed as follows. For all the tested chips, approximately 2.4% of the test chips have large supply currents or a short. The short failure is probably due to the fabrication defects in on-chip decoupling capacitors and will be further discussed in Section 4.4.

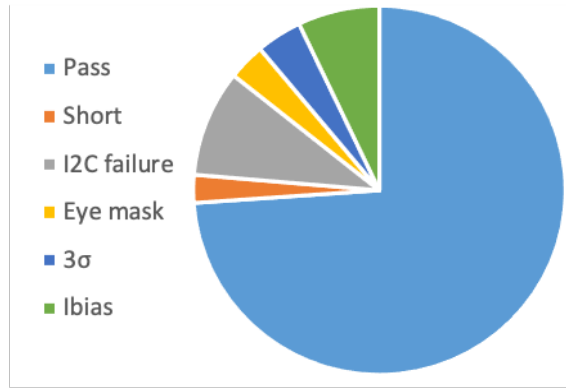

Figure 5. Pie chart of the LOCx2 QC test yield.

About 4.0%, 7.1%, and 3.2% of the tested chips have supply currents beyond three times the standard deviation, a bias current 25% larger or less than the expected value, or fail the eye mask test, respectively. The histograms of the supply current of 3.3 V and 2.5 V and the bias current are shown in figure 6(a-c). These three histograms have similar long tails on their left sides. The reason for the small long tails is still under investigation. Low supply currents of 3.3

V are correlated to low output swings or small modulation currents. Correspondingly, the DAC circuit failure may contribute to the long tail of the measured bias current.

Among all the tested chips, about 9.3% have an I$^2$C communication issue. Our investigation shows that the I$^2$C failure is independent of the I$^2$C operation frequency ranging from 20 kHz to 1 MHz. All the I$^2$C failures follow the pattern where none of the I$^2$C failed chips respond to the external I$^2$C commands. The histogram of the I$^2$C operations that the chips succeeded in is shown in figure 6(d). The maximum number of the I$^2$C operations is 20, which was set in the QC test procedure. About 86.1% of the tested chips passed the I$^2$C test in the first trial, and about 4.5% of the test chips need multiple trials. About 0.14% of chips that need more than ten trails are considered as I$^2$C failures. The percentage of LOCld chips that need multiple I$^2$C operations is almost ten times higher than that of LOCx2 (4.5% for LOCld versus 0.4% for LOCx2).

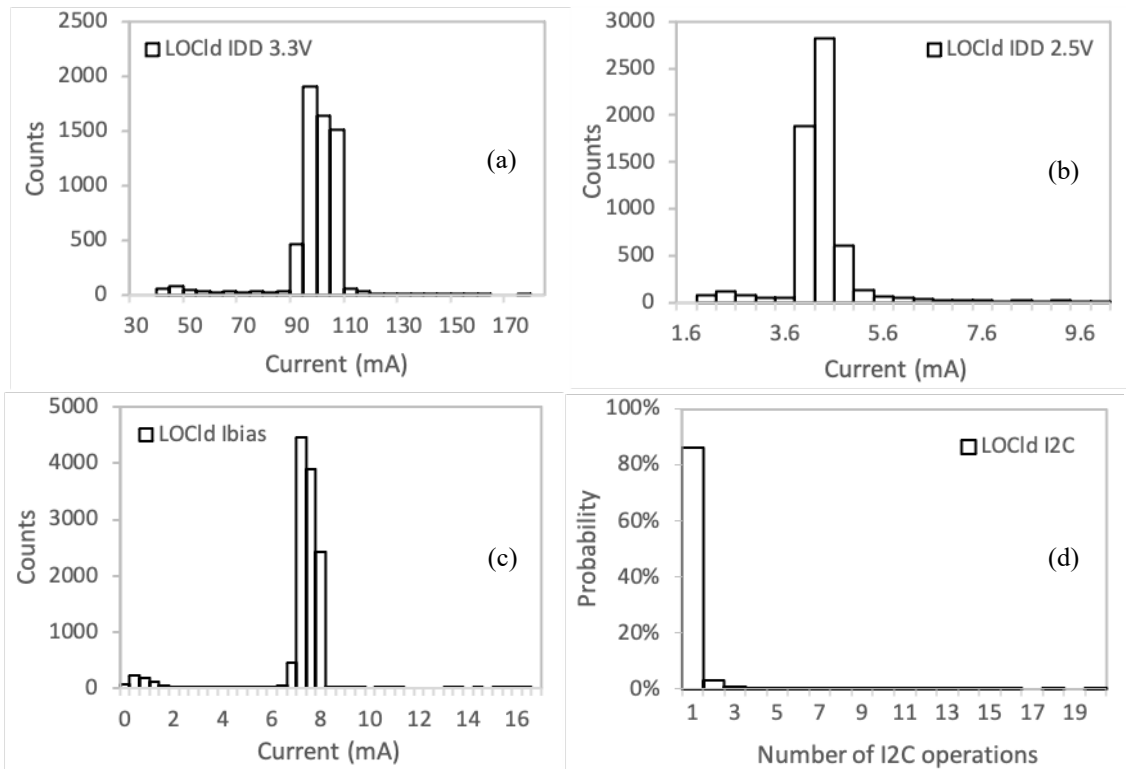

Figure 6. Histograms of the supply current of 3.3 V (a) and 2.5 V (b), the bias current (c), and the number of the I$^2$C operations (d) that eventually succeeded.

Further, LOCld is more sensitive to the signal transition time on the I$^2$C bus than LOCx2, though the I$^2$C target logic in LOCld and LOCx2 are the same synthesized digital functional block. The LTDB tests indicate that the MTx modules located far away from the GBT SCA have a higher failure probability than those near the GBT SCA, whereas such a phenomenon is not true for LOCx2 chips that pair with MTx modules. Our tests reveal that the pull-up resistance of LOCld must be as low as a few hundred ohms, limited by the driving strength of the I$^2$C controller/targets, whereas LOCx2 can tolerate as high as 5 kΩ.

By comparing the design of LOCld with that of LOCx2, we hypothesize that a higher failure rate of LOCld can be attributed to a design issue of LOCld. The SCL and SDA signals in LOCx2 have internal buffers (consisting of two inverters in series) located close to the wire bonding pads. LOCld does not contain internal buffers for these signals. Note that LOCx2 has a much larger dimension than LOCld. The input buffer in LOCx2 is originally intended to drive the trace from the bonding pad to the internal synthesized logic. The input buffer hides the internal load, makes the internal logic less sensitive to the signal transition time, and luckily improves the I$^2$C reliability of LOCx2. LOCld was finalized before LOCx2, and the I$^2$C issue was not realized until the LOCld chips were produced. This is why LOCx2 has the input buffers, but LOCld does not. The hypothesis has been verified in the simulation. The simulation shows that the internal logic simultaneously generates a START signal and a STOP signal when the SDA signal has a slow rise time, which messes up the internal logic. We choose the pull-up resistors on the LTDB to be as low as 316 Ω to reduce the failure rate. We separate the LOCld chips into two grades in the QC test by adjusting the pull-up resistance. The grading will be discussed in Section 3.2.

## 3. QC test of MTx and MTRx

### 3.1 Overview of MTx and MTRx

MTx and MTRx are dual-channel optical transmitter (MTx) and optical transceiver (MTRx) modules. MTx has two Transmitter Optical Sub-Assembly (TOSAs), while MTRx has a TOSA plus a ROSA. MTx and MTRx modules use standard hermetically packaged TOSAs/ROSAs with Lucent Connector (LC) ferrule receptacles to ensure light coupling efficiency. The outer dimension of the cross-section of the TOSA/ROSA is 5.9 mm. The TOSA is a Commercial-Off-The-Shelf (COTS) part produced by Truelight (part number TTF-1F59-427). The ROSA is a custom part produced by CERN with a P-I-N diode and GBTIA.

The most critical mechanical design element of MTx/MTRx is a custom latch. The latch couples two fibers with LC ferrules, flanges, and springs to the TOSA/ROSA and attaches the TOSA/ROSA to the module PCB. The latch has two parts. The first part anchors the TOSA/ROSA to the module PCB with a screw through a hole in the center beam and MTx/MTRx to the motherboard, while the second part pushes the LC ferrules on the fibers into the TOSA/ROSA through the springs. The latch is injection molded with Ultem 1000 PEI, the same material as TOSA and ROSA bodies. The fibers are ordered from manufacturers that assemble standard LC connectors onto fibers.

The carrier PCB is 0.9 mm thick and has two cutouts to accommodate the TOSA/ROSA so that MTx and MTRx are as thick as the outer dimension of the TOSA/ROSA. The electrical interface is a high-speed hermaphroditic connector produced by Samtec (part number LSHM-120-02.5-L-DV-A-N-TR) with a 5.0 mm stacking height. MTx/MTRx is pluggable and mid-board mountable. The dual-channel serializer LOCx2 can be installed under MTx. On the top of the module is a 0.1 mm thick thermal tape.

As the communications from the LTDB to the off-detector electronics in the counting room must be established at power-on for configuration purposes, MTRx must be functional with the default configuration values in LOCld. The power-on reset circuit is implemented in MTx and MTRx. The power-on reset circuit is connected to the active-low reset pin of LOCld and comprises a 10 kΩ pull-up resistor and a 22 μF capacitor connected to the ground.

The photographs of a standalone latch part 1, a complete latch, an MTRx with the top LOCld, and an MTRx with the top thermal tape are shown in figure 7.

### 3.2 Procedure of MTx/MTRx QC test

The QC test procedure of MTx/MTRx is similar to that of the LOCld QC test procedure. First, the eye mask test of each MTx or MTRx was checked after power-on-reset without configuration at 5.12 Gbps. Any module that failed the eye mask test was considered an eye mask failure. Second, the module was configured via I²C. Any module with the read-back values inconsistent with the written values was considered an I²C failure. Third, the eye diagram of the module after configuration was checked. Eye mask parameters, including Average Optical Power (AOP), Optical Modulation Amplitude (OMA), Extension Ratio in dB (EXdB), rise time, fall time, peak-to-peak jitter (PPJ), and RMS jitter (RMSJ), and the supply currents were measured. Finally, the modules that failed the QC test were reworked. The acceptance criteria of the QC test are listed in Table 2. For the receiving channel of the MTRx module, only the eye diagram was recorded.

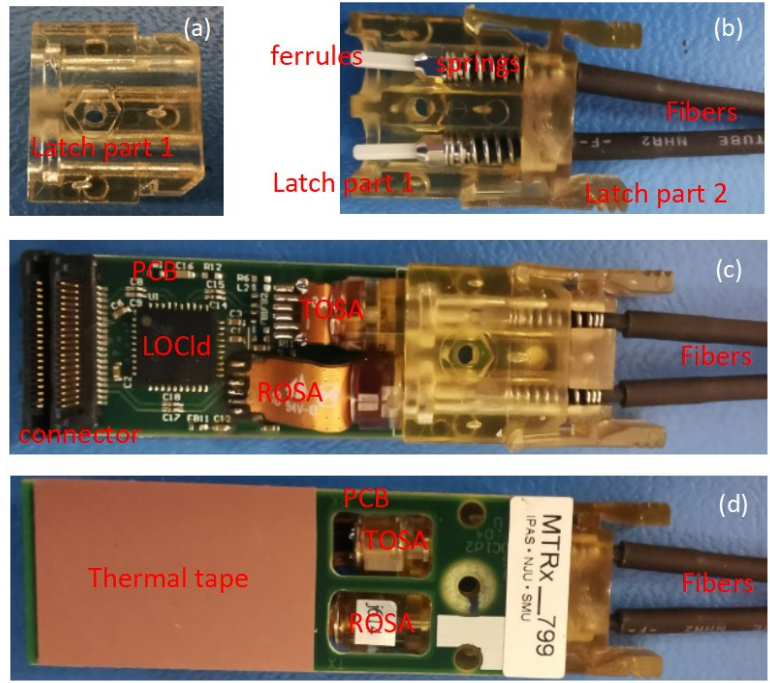

Figure 7. Photographs of a standalone latch part 1 (a), a complete latch with fibers plugged in (b), an MTRx with the top LOCld and the fibers plugged in (c), and an MTRx with the top thermal tape and fibers plugged in.

As discussed in Section 2.4, LOCld is sensitive to the signal transition time on the I²C bus. To work around such an issue, we added an extra step to separate each MTx module into two grades. Grade A can tolerance a longer transition time than Grade B. The threshold to separate two grades is determined by allowing each grade to have roughly the same number of modules. The threshold is implemented in the QC test by adjusting the pull-up resistance. MTRx modules are not graded because they are located close to the GBT SCA.

Table 2: Acceptance criteria of the MTx QC test.

| Parameters | Acceptance criteria | | Unit |
|---|---|---|---|
| | Min | Max | |
| AOP | -3.5 | | dBm |
| OMA | 0.3 | | mW |
| EXdB | 3 | | dB |
| Rise time | | 80 | ps |
| Fall time | | 80 | ps |
| PPJ | | 35 | ps |

## 3.3 Setup of MTx/MTRx QC test

The block diagram of the MTx/MTRx QC test setup is shown in figure 8(a-b). A pattern generator produced by Centellax (Model number PCB12500 with a clock synthesizer TG1C1-A) generated a PRBS pattern $2^7$-1. Since MTx has dual channels, we used an evaluation board produced by Analog Devices, Inc. (part number 125614-HMC850LC3) to provide a fanout. We utilized a sampling oscilloscope (Tektronix, Model number TDS80000B with an optical sampling probe 80C012) to monitor eye diagrams. The pattern generator also generated a synchronous clock for the sampling oscilloscope as the trigger. For MTRx, the optical output of the transmitter was looped back to feed the optical receiver. The optical sampling oscilloscope was replaced by a real-time electrical oscilloscope (Tektronix, Model TDS72004A) with a second synchronous clock from the pattern generator as the trigger. The chip under test was configured via a USB-to-I$^2$C adapter (Robot Electronics USB-ISS Multifunction USB Communication Module). For both MTx and MTRx, a laptop running a LabVIEW program automatically controlled the test procedure. A photograph of the test setup is shown in figure 8(c).

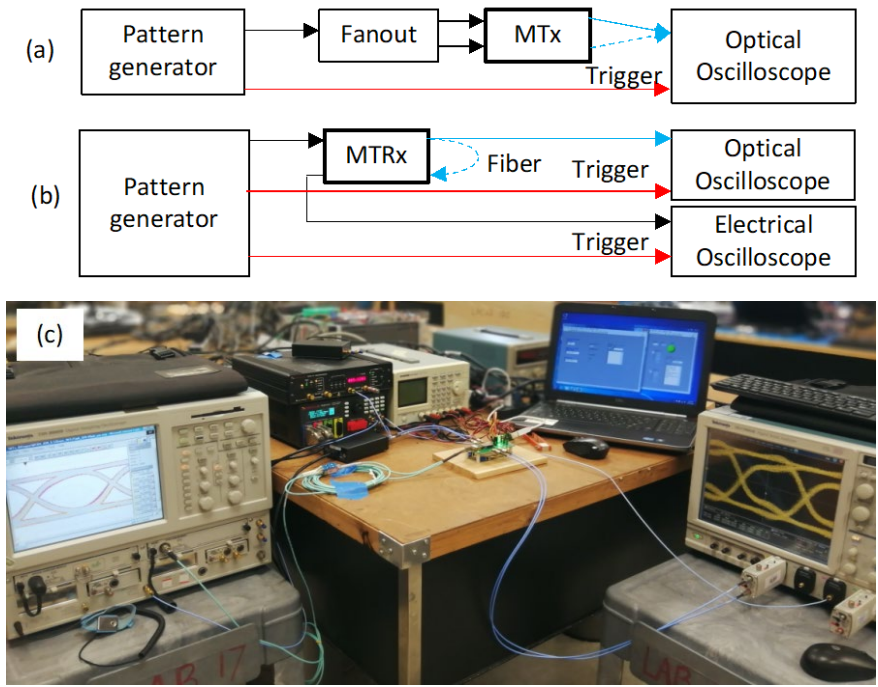

Figure 8. Block diagrams of the MTx (a) and MTRx (b) test setup and photograph (c) of the MTRx test setup.

## 3.4 Results of MTx/MTRx QC test

In the QC test, 3275 out of 3343 MTx modules and 797 out of 810 MTRx modules passed the QC test. The yields are 98.0% and 98.4% for the MTx and MTRx, respectively. Three MTRx modules failed in the initial test and passed the test after TOSAs or ROSAs were reworked (not included in the yield listed above). Thirty MTRx modules are being assembled and will be tested. All failed MTx modules are to be reworked. Figure 9 is a screenshot of a typical eye diagram with masks. If any sample points fall inside the prohibition regions, the module is considered to fail the test. The histograms of the AOP and the OMA of all transmitter channels (including MTRx modules) are displayed in figure 10. The AOP and the OMA correspond to the bias and modulation currents of LOCld, respectively.

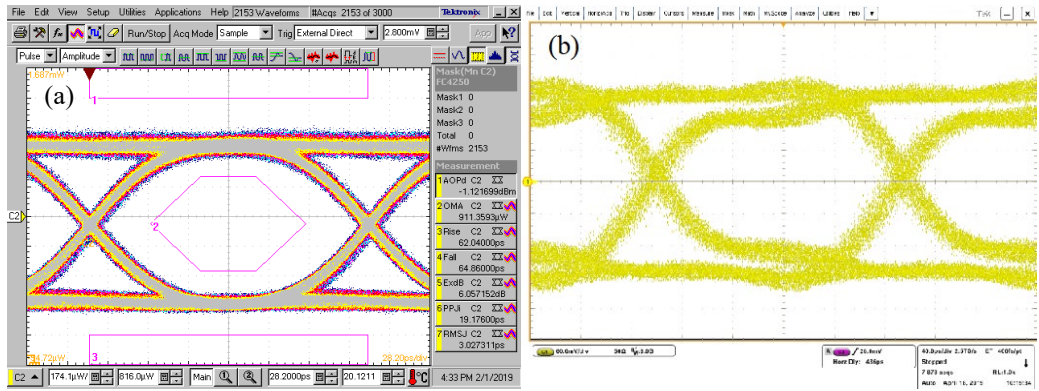

Figure 9. Typical eye diagrams of an MTx and an MTRx receiver channel.

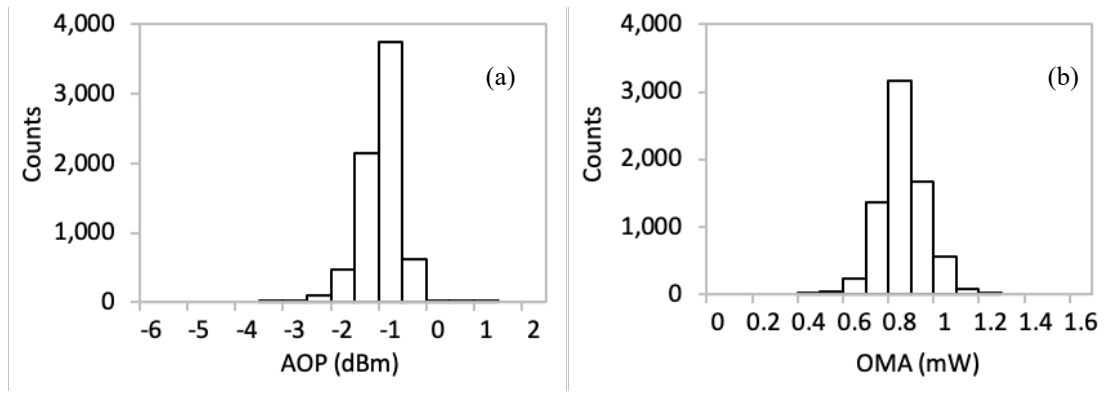

Figure 10. Histograms of the AOP and the OMA of all transmitter channels.

## 4. QC of LOCx2

### 4.1 Overview of LOCx2

LOCx2 is a dual-channel serializer ASIC designed for the optical data link to read out the LTDB. The block diagram of LOCx2 and its interfaces are shown in figure 11(a). LOCx2 includes a Phase-Locked Loop (PLL), two encoders, two 16:1 serializers, two Current Mode Logic (CML) line drivers, and an $I^2C$ target. The PLL has an inductor-capacitor-tank-based voltage-controlled oscillator with digitally selectable four tuning bands (Bands 0-3). The PLL requires a reference clock of about 40 MHz provided by the GBTX and generates an output clock of 2.56 GHz and other clocks required by the serializers and encoders. Each LOCx2 transmits the data from four quad-analog-channel ASIC ADC chips [13] or two octal-analog-channel ASIC ADC chips (Texas Instruments, Part number ADS5272 or ADS5294) [14]. The encoder implements a custom line code [15]. Each serializer takes the digitized data of eight analog signals. Each serializer has 16-bit parallel data inputs and serial data output and consists of four stages of 2:1 multiplexers in a binary-tree structure. The CML line driver is composed of five-stage CML differential amplifiers. The serial output is CML with a minimum swing of 2 mA.

The dimension of the LOCx2 die is 3.86 mm by 6.036 mm. LOCx2 is packaged in a 100-pin 0.4-mm-pitch QFN package. The chip dimension is 12 mm by 12 mm. In the QC test, roughly 7,200 LOCx2 chips were screened. Packaged chips were shipped in tubes. During the tests, the packaged chips were handled with vacuum pens. Chips that passed the QC test were put onto the storage trays and stored in a low vacuum with a desiccant. The chips that failed the QC test were grouped into different failure modes for further investigations. A photograph of package LOCx2 chips on a storage tray is shown in figure 11(b).

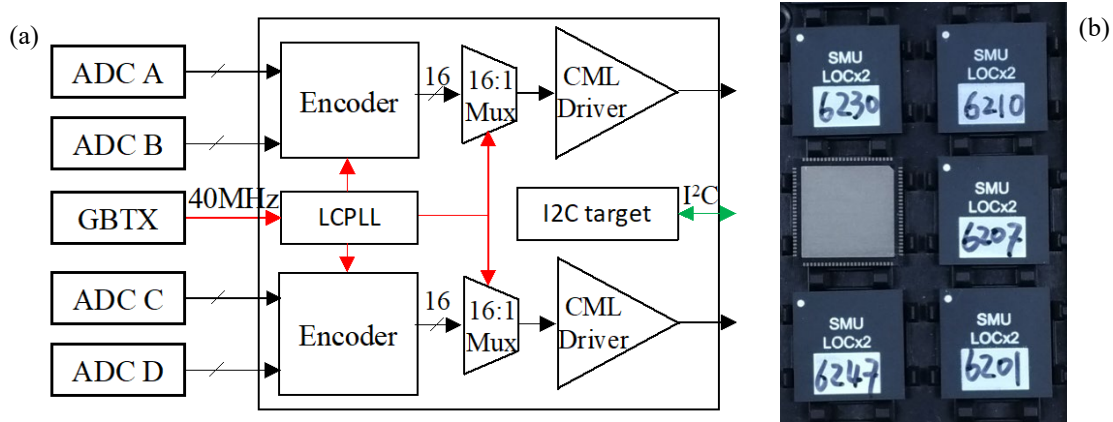

Figure 11. Block diagram (a) and photograph (b) of LOCx2.

**4.2 Procedure of LOCx2 QC test**

The QC test includes two stages. The first stage is the eye diagram test. First, the chip was supplied with the nominal voltage of 2.5 V. If the chip had a supply current higher than 600 mA, which was about twice the nominal value, it was removed as a short failure. If the chip had a current below 250 mA, which was about 1/3 less than the nominal value, it was removed as a low current failure. Second, the chip was configured to be in the test mode, in which pseudo-random test-pattern data were generated internally. The register values were read back and compared with the written values. If the read values and the written values were not consistent, the chip was removed as an $I^2C$ failure. Third, the eye mask was checked, and the chip with eye mask failure was removed. Lastly, the test clock output signal was measured. The eye diagram and the test clock output signal were checked at the reference clock of 39.4, 40.0, and 40.4 MHz and the PLL Bands of 0 and 1. Note that PLL Bands 2-3 were not swept because they could not cover these three input frequencies.

The second stage is the BER test. When the input data were in the ASIC ADC mode, the BER was measured at the nominal supply voltage of 2.50 V for 15 minutes, 2.25 V for 1.5 minutes, and 2.75 V for 1.5 minutes, respectively. No error in 15 minutes and 1.5 minutes corresponds to a BER of $1\times10^{-12}$ and $1\times10^{-11}$ at the confidence level of 95%, respectively. A chip with an error in this test was removed as a BER failure. Then the chip was configured to be in the ADS5272 mode, and the BER was measured at 2.5 V for 1.5 minutes. Once the chip was determined to have errors, the QC test was stopped instead of waiting for the entire test time.

**4.3 Setup of LOCx2 QC test**

The test setup of the eye diagram test is shown in figure 12. An evaluation board (Silicon Lab, part number Si5338EVB) generated a reference clock at 39.4, 40.0, or 40.4 MHz and was fed to LOCx2. Each chip was placed in a socket (Plastronics, part number 100QHC40A21212) on a LOCx2 carrier board. The chip was configured via a USB-to-$I^2C$ adapter (Robot Electronics USB-ISS Multifunction USB Communication Module) to be in the test mode, and pseudo-random data were generated internally. Each serial output was connected to a differential probe of a real-time oscilloscope. The test clock output of LOCx2 was observed to determine if the PLL was locked to the reference clock.

The block diagram and the photograph of the BER test setup are shown in figure 13. A Xilinx Kintex-7 Field Programmable Gate Array (FPGA) evaluation board (part number

KC705) generated 24 differential pairs of data (the thick green lines in figure 13(a)) required by a single LOCx2 chip. A High-Pin-Count (HPC) FPGA Mezzanine Card (FMC) fanned out the data to six carrier boards, each with a socket, so that six LOCx2 chips were measured simultaneously. Fanout chips CDCLVD1212 produced by Texas Instruments feature a channel skew of less than 600 ps. The data generated in the FPGA have an adjustable delay with a resolution of 80 ps. The phase skew among different input data was measured to be within 400 ps (peak-to-peak), within the specifications of LOCx2.

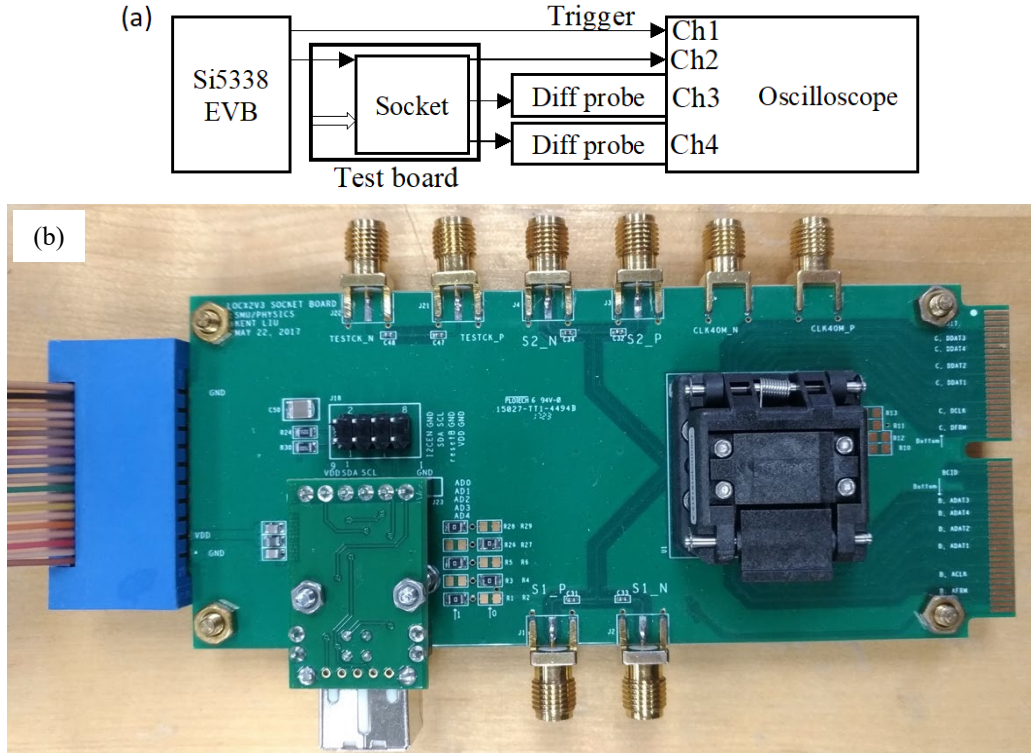

Figure 12. Block diagram (a) and photograph (b) of the QC test setup.

The FPGA had 12 multi-gigabit transceivers to receive 12 serial data (the blue lines in figure 13(a)) from the six chips via Sub-Miniature A (SMA) coaxial cables and checked if the serial data had any errors. The rate of the serial data was 5.12 Gbps per channel. The mezzanine card and a Peripheral Component Interconnect Express (PCIe) extension board connected two and four LOCx2 chips to the FPGA.

Three clocks were generated on an Si5338 evaluation board. CLK0 and CLK1 were for the FPGA pseudo-random data generation and the 12 FPGA transceivers, respectively. CLK2 was fanned out to the six LOCx2 chips by a clock generator chip (Texas Instruments, part number LMK03200).

The QC procedure was controlled automatically in a LabVIEW program running on a laptop. The serial numbers were prompted to input after the chips were placed in sockets. The LOCx2 chips under test were configured via an I$^2$C interface module (National Instruments, part number USB-8451). The I$^2$C signals were fanned out on the mezzanine card. During the test, the internal registers of LOCx2 were written and then read back. If the values read back were not consistent with the written values, an I$^2$C error was recorded.

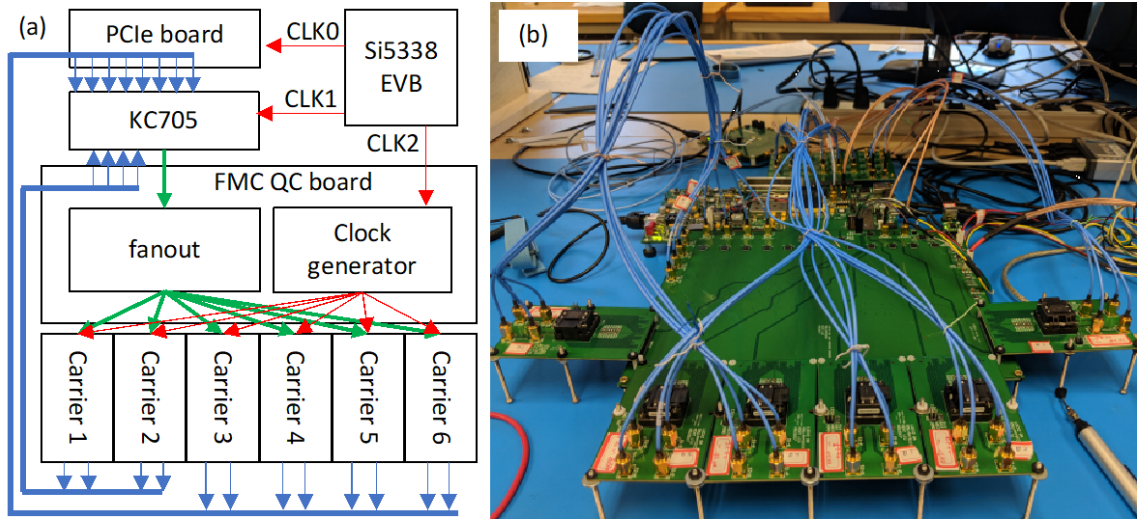

Figure 13. Block diagram and photograph of the BER test setup.

### 4.4 Results of LOCx2 QC test

A total of 5163 LOCx2 chips were packaged in production packaging batches, and 3198 chips passed the full QC test. The overall yield of the QC test is 61.9%. The eye diagram and the typical test clock output of a sample chip in production packaging batches are shown in figure 14. In addition to the production packaging batches, 424 chips out of about 2000 chips in prototype packaging batches passed the full QC test. The chips in prototype packaging batches will only be used as spares. The yield of the chips of the prototype packaging batches is very low because of immature packaging technologies. The following QC test results are all based on the chips in the production packaging batches.

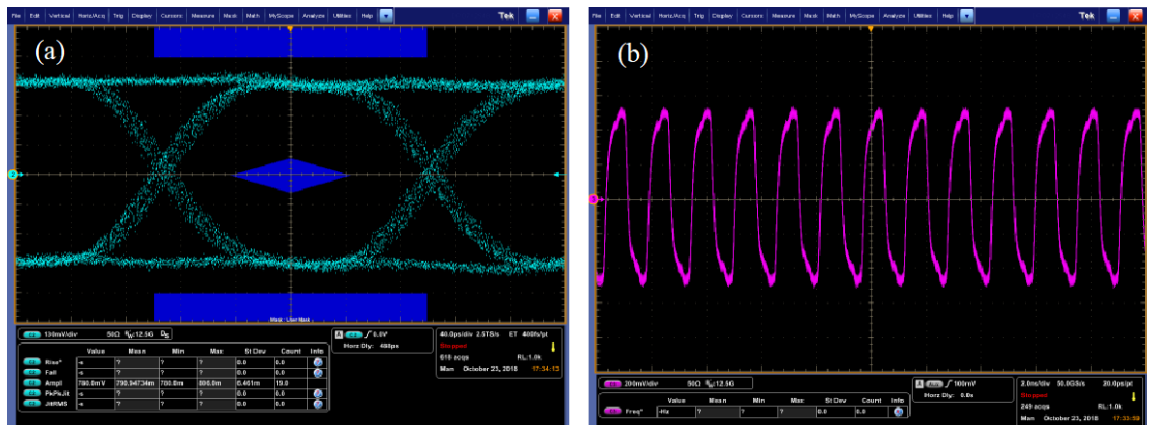

Figure 14. Eye diagram and test clock output of a sample LOCx2 chip.

The percentage of each failure mode of all production packaged chips of LOCx2 is shown in figure 15. These failure modes represent errors occurring in the encoders, the I$^2$C target, decoupling capacitors, the serializers/output drivers, and the other blocks. The QC test yield of LOCx2 is lower than that of LOCld (61.9% versus 73.9%) because LOCx2 is more complicated than LOCld. The same characteristic explains the percentage of eye mask failure (8.6% in LOCx2 versus 3.2% in LOCld). Any problem in the serializer or the output driver in each channel of LOCx2 contributes to the eye mask failure, whereas only the amplifier and output driver in each channel of LOCld are related to the eye mask failure.

The percentage of shorts in LOCx2 is much higher than that of LOCld (9.8% versus 2.4%). The failure does not seem to be caused by electrical static discharge (ESD) in the packaging or in the QC procedures, where strict ESD protection rules are followed. Moreover, in both LOCld and LOCx2, all input/output pins are protected with custom-designed ESD protection diodes. The shorts are probably due to the defects in Metal-Insulator-Metal (MIM) capacitors used in the decoupling capacitors in fabrication. LOCx2 uses a large area of decoupling capacitors implemented in regular NMOS transistors and MIM capacitors. The process that LOCx2 is fabricated in does not have an upper limit of the MIM capacitor area to preserve the chip yield as many mainstream commercial CMOS processes do. This hypothesis is supported by the fact that LOCx2 has much more decoupling capacitance than LOCld. This hypothesis is also verified by the correlation between the decoupling capacitance and the number of shorts. LOCx2 has six separate power supply rails, each with a different quantity of decoupling capacitance. We selected 143 shorted chips and found out which power-supply rail was shorted. The number of shorts of each power supply rail was counted and compared with the corresponding decoupling capacitance. The correlation between the percentage of shorts in the six power-supply rails and the decoupling capacitance is shown in figure 16. As shown in figure 16, the percentage of shorts is roughly proportional to the decoupling capacitance.

The histogram of the supply current is shown in Figure 17(a). The average supply current is 331.6 mA, and the standard deviation is 10.5 mA.

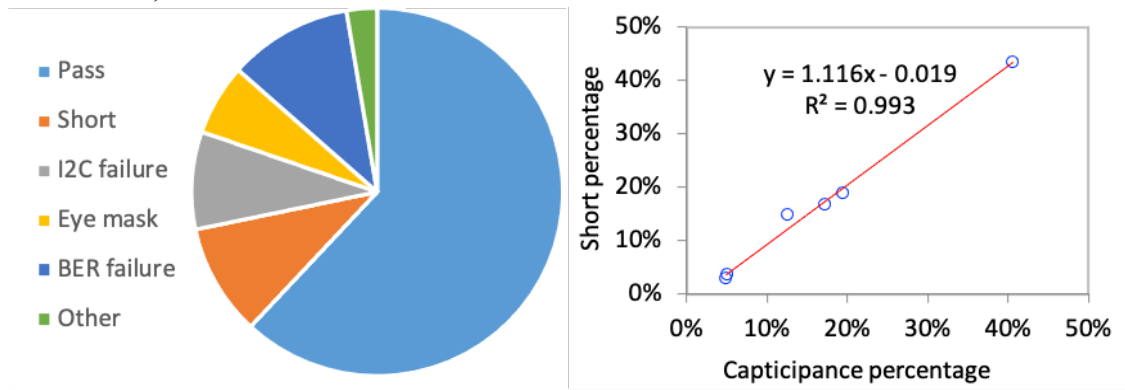

Figure 15. Pie chart of the LOCx2 QC yield.   Figure 16. Capacitance versus shorts.

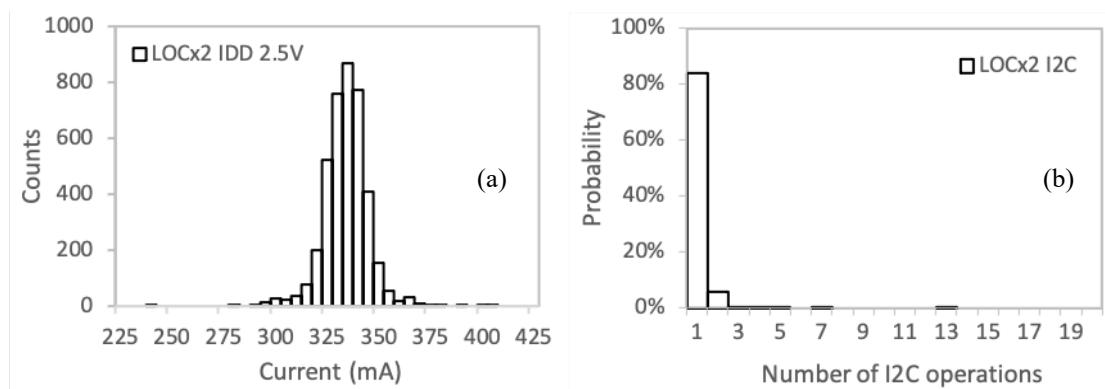

Figure 17. Histograms of the supply current of 2.5 V (a) and the number of the I²C operation that finally succeeded (b).

The percentage of the I²C communication error in LOCld is comparable with that of LOCx2 (9.8% for LOCx2 versus 9.4% for LOCld). The histogram of the I²C operation numbers that passed the I²C test in LOCx2 is shown in figure 17(b). The maximum number of the I²C operations is 20, which is set in the QC test procedure. For LOCx2, 84.1% of the chips succeeded in the first trial, and 5.7% succeeded in the second trial. For LOCld, 86.1% of the chips succeeded in the first trial, and 3.0% succeeded in the second trial. The minor percentage difference of the chips that succeeded in the first trial in LOCx2 and LOCld is explicable. The LOCx2 test board has no power-on reset circuit, whereas the LOCld test board does. The $I^2C$ target functional block of LOCx2/LOCld may fail before a reset, a write, or read operation. The percentage that LOCx2 succeeded in the first two trials is close to that of LOCld (89.8% for LOCx2 versus 89.1% for LOCld). A small percentage of chips needed more than two trials (0.4% for LOCx2 and 1.4% for LOCld). The percentage difference has been explained in Section 2.4. In the QC test, about 0.02% of LOCx2 chips and 0.1% of LOCld chips need more than ten trials to succeed and are considered to fail as an $I^2C$ failure.

## 5. Conclusion

The QC tests of the front-end optical link components, including LOCld, MTx, MTRx, and LOCx2, are conducted for the ATLAS Liquid Argon Calorimeter Phase-1 upgrade. A total of 5341 LOCld chips, 3275 MTx modules, 797 MTRx modules, and 3198 LOCx2 chips are qualified. The yields are 73.9%, 98.0%, 98.4%, and 61.9%, for LOCld, MTx, MTRx, and LOCx2. Besides, 424 LOCx2 chips in prototype packaging batches pass the full QC tests and will only be used as spares.

## Acknowledgments

This work is supported by the US-ATLAS R&D program for LHC upgrade, SMU's Dedman Dean's Research Council Grant, and the National Science Council in Taiwan. We are also grateful to the Hamilton family for their contributions to support undergraduate research at SMU.